\documentclass[preprint,aps,prc,showpacs,nofootinbib]{revtex4-1}
\usepackage{epsfig}
\usepackage{graphicx,amsmath}

{

\newcommand\nn{\nonumber}

\def\phi{\varphi}

\newcommand\ba{\begin{eqnarray}}
\newcommand\ea{\end{eqnarray}}
\newcommand\be{\begin{equation}}
\newcommand\ee{\end{equation}}

\begin{document}
\title{General analysis and numerical estimations of polarization
observables in $\bar N+N\to \pi+ e^++e^-$ reaction in an exclusive
experimental setup}

\author{G. I. Gakh}
\affiliation{\it National Science Centre "Kharkov Institute of Physics and Technology"\\ 61108 Akademicheskaya 1, Kharkov,
Ukraine }
\author{E.~Tomasi-Gustafsson}
\email[E-mail: ]{etomasi@cea.fr}
\altaffiliation{Permanent address: \it CEA,IRFU,SPhN, Saclay, 91191 Gif-sur-Yvette Cedex, France}
\affiliation{\it Univ Paris-Sud, CNRS/IN2P3, Institut de Physique Nucl\'eaire, UMR 8608, 91405 Orsay, France}
\author{A. Dbeyssi}
\affiliation{\it Univ Paris-Sud, CNRS/IN2P3, Institut de Physique Nucl\'eaire, UMR 8608, 91405 Orsay, France}
\author{A. G. Gakh}
\affiliation{Kharkov National University, 61077 4 Svobody Sq., Kharkov, Ukraine}
\begin{abstract}
The dependence of the differential cross section and polarization
observables in the $\bar N+N\to \pi +e^++e^-$ reaction on the
polarizations of the proton target and antiproton beam (the produced
electron may be unpolarized or longitudinally polarized) have been
derived in a general form using hadron electromagnetic current
conservation and $P-$ invariance of the hadron electromagnetic
interaction. The analysis was done for the case of an exclusive
experimental setup where the produced electron and pion are detected in
coincidence. The explicit dependence of all polarization observables
on two, from five, kinematic variables (the azimuthal angle
$\varphi$ and  the virtual-photon parameter $\epsilon$), have been
obtained assuming one-photon-exchange.
The application to the particular case of a mechanism which contains
information on time-like proton form factors in the unphysical region is 
considered in the Born approximation.
\end{abstract}

\maketitle

\section{Introduction}
Renewed interest in the reactions induced by antiprotons is related
to the possibility to accelerate high intensity antiproton beams up
to 15 GeV/c momentum at the FAIR facility \cite{FAIR}. The wide
program foreseen at this facility by the  PANDA collaboration
\cite{PANDA} is focused on strong interaction and includes
charmonium spectroscopy, hybrids, glueballs, and charm in nuclei (for a
review, see \cite{Wiedner:2011mf}). Moreover, the selection in the final state of
a pair of leptons accompanied by a pion will allow to study aspects of the
electromagnetic interaction. In particular, it allows to access
electromagnetic proton form factors in the so-called 'unphysical
region'. Assuming the dominance of $t$
and $u$ channel exchange diagrams \cite{Ad07}, the emission of a
pion by the proton or the antiproton lowers the momentum 
squared ($q^2$) of a virtual photon which subsequently decays into a
lepton pair. This mechanism allows to reach $q^2$ values under the
$4M^2$ threshold ($M$ is the proton mass) in the time-like region of
transferred momenta.

This idea was firstly suggested by M.P. Rekalo, using the crossing
symmetry related reaction $\pi +N\to N+e^-+e^+$ \cite{MPR,DDR}. Pion
scattering on nucleon and nuclei will be investigated in the
resonance region by the HADES collaboration with the pion beam available at GSI (Darmstadt) in
the next future \cite{Hades}. The third reaction related by crossing
symmetry to the previous ones is the pion electroproduction on proton,
which has been widely investigated in semi-inclusive and exclusive
polarized and unpolarized experiments, in different kinematical
regions.

In this paper we extend our previous analysis \cite{Gakh:2010qp} for
the case of exclusive measurements, in an experimental setup where
all final particles are detected. The five-fold differential cross
section is derived in unpolarized and polarized case. Different
polarization phenomena are considered with polarized or unpolarized beam and
target, when the polarization of the scattered electron is measured.  
The expressions of the observables are given in terms of the
six independent amplitudes (in general complex functions of three
kinematical variables) which fully describe the considered reactions
in the one-photon exchange approximation. A numerical application is
done for the model suggested in Ref. \cite{Ad07}.

\section{General formalism}

\subsection{General expression of the cross section}
The general structure of the differential cross section for the
reaction: 
\be \bar p(p_1)+p(p_2)\to \gamma^*+\pi^0 \to
\ell^+(k_2)+\ell^-(k_1)+\pi^0(q_{\pi}),~\ell=e,\mu,~\tau
\label{eq:eq1} 
\ee 
is considered in the frame of one-photon-exchange
mechanism, and applied to the case of electrons, neglecting the lepton mass.

In this section, the formalism is based on the most general symmetry properties of the
hadron electromagnetic interaction, such as gauge invariance
(the conservation of the hadronic and leptonic electromagnetic
currents), P-invariance (invariance with respect to space
reflections) and does not depend on the details of the reaction
mechanism.

Let us consider the reaction $\bar p+p\to\gamma^* +\pi^0 $
where $\gamma^*$ is a virtual
photon. In the one-photon-exchange approximation, the matrix
element of the reaction (\ref{eq:eq1}) in terms of the hadronic $J_{\mu}$ and
leptonic $j_{\mu}$ currents can be written as
\be
{\cal M}=\frac{4\pi\alpha}{q^2}J_{\mu}j_{\mu},
\label{eq:eqM}
\ee
where $q=k_1+k_2$ is the virtual photon four-momentum, $J_{\mu}$ is
the electromagnetic current describing the transition  $\bar p+p\to
\pi^0+\gamma^* $,  $j_{\mu}=\bar
u(-k_2)\gamma_{\mu}u(k_1)$ describes the decay  $\gamma^*(q)\to e^+(k_2)+e^-(k_1)$.

The modulus squared of the matrix element can be written as the
contraction of the hadronic $H_{\mu\nu}$ and leptonic  $L_{\mu\nu}$ tensors:
\be
|{\cal M}|^2=\frac{16\pi^2\alpha^2}{q^4}H_{\mu\nu}L_{\mu\nu},
\label{eq:eqM2}
\ee
with
\be
H_{\mu\nu}=J_{\mu}J_{\nu}^*, \ \    L_{\mu\nu}=j_{\mu}j_{\nu}^*.
\label{eq:eq4}
\ee
The most convenient system for the analysis of the spin structure of the
matrix element is the antiproton-proton center-of-mass system (CMS).

To derive polarization observables it is necessary to define a
particular reference frame. We choose a reference frame with the $z$
axis directed along the momentum of the virtual photon ${\vec k}$,
the momentum of the antiproton beam ${\vec p}$ lies in the $xz$
plane. The $y$ axis is normal to the $\bar p+p\to \pi^0+\gamma^*$
reaction plane and it is directed along the vector ${\vec k}\times
{\vec p}$; $x$, $y$, and $z$ form a right-handed coordinate system.

In this system, the components of the particle four-momenta for the
reaction $\bar p+p\to\pi^0+\gamma^*$ are
\be
p_1=(E, {\vec p}), \ p_2=(E, -{\vec p}), \ q=(k_0, {\vec k}),
\ q_{\pi}=(E_{\pi}, -{\vec k}).
\label{eq:eqmom}
\ee
The general expression for the differential cross section of
the reaction considered has a standard form
\be
d\sigma =\frac{\alpha^2}{(2\pi
)^3}\frac{L_{\mu\nu}H_{\mu\nu}}{Iq^4}\frac{d^3k_1}{2E_1}\frac{d^3k_2}{2E_2}
\frac{d^3q_{\pi}}{2E_{\pi}}\delta^{(4)}(p_1+p_2-k_1-k_2-q_{\pi}),
\label{eq:eqsigma}
\ee
where $I^2=(p_1\cdot p_2)^2-M^4, $ $M$ is the nucleon mass,
$E_1(E_2)$ is the energy of the electron (positron) and $E_{\pi}$ is
the pion energy.

Integrating the expression (\ref{eq:eqsigma}) over the positron variables and over the electron energy, with the help of  $\delta -$function, we
obtain the following form for the differential cross section
(hereafter we neglect the electron mass)
\be
\frac{d^3\sigma}{dE_{\pi}d\Omega_{\pi}d\Omega_e}
=\frac{\alpha^2}{64\pi^3}\frac{1}{q^4}\frac{E_1|{\vec
k}|}{pW}\frac{L_{\mu\nu}H_{\mu\nu}}{k_0-|{\vec k}|\cos\theta_1},
\label{eq:eqsigma3}
\ee
where $W=2E$ is the total energy of the final particles,
$p=\sqrt{W^2-4M^2}/2$ is the modulus of the three-momentum of the proton or antiproton in
the reaction CMS, $E_1$ is the energy of
the electron, $\theta_1$ is the  angle between the momenta of the virtual photon and of the detected electron,  $k_0=2E-E_{\pi}$  and $|{\vec
k}|=\sqrt{E_{\pi}^2-m_{\pi}^2}$ are the energy  and the momentum of the
virtual photon, and $m_{\pi}$ is the pion mass. The electron energy
in the reaction CMS is $E_1=q^2/2(k_0-|{\vec k}|\cos\theta_1)$, with
$q^2=4E^2-4EE_{\pi}+m_{\pi}^2$.

The leptonic tensor, when the electron is unpolarized, is:
\be
L_{\mu\nu}(0)=-2q^2g_{\mu\nu}+4(k_{1\mu}k_{2\nu}+k_{1\nu}k_{2\mu}).
\label{eqL}
\ee
The longitudinal polarization of the electron induces a term in the
leptonic tensor which has the form
\be
L_{\mu\nu}(s)=2i\lambda <\mu\nu qk_1>,
\label{eqHL}
\ee
where $<\mu\nu ab>=\varepsilon_{\mu\nu\rho\sigma}a_{\rho}b_{\sigma}$, and $\lambda $ is
the degree of the longitudinal polarization of the electron. We use
the convention $\varepsilon_{xyz0}=1$.

Taking into account the conservation of the hadronic $J_{\mu}$ and
leptonic $j_{\mu}$ currents, we can remove the time components of
the hadronic and leptonic tensors and obtain:
\ba
L_{\mu\nu}H_{\mu\nu}&=&\frac{1}{2}(L_{xx}+L_{yy})(H_{xx}+H_{yy})+
\frac{1}{2}(L_{xx}-L_{yy})(H_{xx}-H_{yy})+\frac{q^4}{k_0^4}L_{zz}H_{zz}+
\nn\\
&&
\frac{1}{2}(L_{xy}+L_{yx})(H_{xy}+H_{yx})+\frac{1}{2}(L_{xy}-L_{yx})
(H_{xy}-H_{yx})+ \nn\\
&&\frac{q^2}{2k_0^2}(L_{xz}+L_{zx})(H_{xz}+H_{zx})
+\frac{q^2}{2k_0^2}(L_{xz}-L_{zx})(H_{xz}-H_{zx})+\nn\\
&&
\frac{q^2}{2k_0^2}(L_{yz}+L_{zy})(H_{yz}+H_{zy})+
\frac{q^2}{2k_0^2}(L_{yz}-L_{zy})(H_{yz}-H_{zy}).
\label{eq:HL}
\ea
 Let us express
the components of the leptonic tensor in terms of measured
quantities. The following relations, which can be derived from the
energy and momentum conservation in our coordinate system, hold:
\ba
k_{1x}&=&E_1\sin\theta_1\cos\phi , \ k_{1y}=E_1\sin\theta_1\sin\phi
, \
k_{1z}=E_1\cos\theta_1,\nn\\
k_{2x}&=&-k_{1x}, \ k_{2y}=-k_{1y}, \
k_{2z}=|{\vec k}|-k_{1z},
\label{eq:ki}
\ea
 where $\phi $ is the azimuthal angle of
the electron momentum, i.e., the angle between the $xz$ plane and
the electron-positron pair production plane which is determined by
the momenta of the virtual photon and detected electron.

Then the general structure of the differential cross section for the
reaction (\ref{eq:eq1}) for the case when produced electron and the pion are
detected in coincidence and the detected electron is longitudinally
polarized (the polarization states of the proton target and
antiproton beam can be any) has the form
\ba
\frac{d^3\sigma}{dE_{\pi}d\Omega_{\pi}d\Omega_e}
&&=N\bigl\{H_{xx}+H_{yy}+\varepsilon \cos2\phi
(H_{yy}-H_{xx})+2\varepsilon \frac{q^2}{k_0^2}H_{zz}-\varepsilon
\sin2\phi (H_{xy}+H_{yx})-
\nn\\
&&
\frac{\sqrt{q^2}}{k_0}\sqrt{2\varepsilon (1-\varepsilon )}\eta \cos\phi
(H_{xz}+H_{zx})-\frac{\sqrt{q^2}}{k_0}\sqrt{2\varepsilon
(1-\varepsilon )}\eta \sin\phi (H_{yz}+H_{zy})- \nn\\
&&
i\lambda \frac{\sqrt{q^2}}{k_0}\sqrt{2\varepsilon (1+\varepsilon )}\sin\phi
(H_{xz}-H_{zx})+\nn\\
&&i\lambda \frac{\sqrt{q^2}}{k_0}\sqrt{2\varepsilon
(1+\varepsilon )}\cos\phi (H_{yz}-H_{zy})
+i\lambda \sqrt{1-\varepsilon^2)}\eta (H_{xy}-H_{yx})\bigr\}, ~\nn\\
&& N=\frac{\alpha^2}{32\pi^3}\frac{1}{q^4}\frac{E_1^2|{\vec
k}|}{pW}\frac{1}{1+\varepsilon }, 
\label{eq:eqd3s} 
\ea 
where $\eta=sign(E_1-E_2)$. Taking into account that $E_1-E_2=|{\vec
k}|(k_0\cos\theta_1-|{\vec k}|)/(k_0-|{\vec k}|\cos\theta_1)$ and
$k_0-|{\vec k}|\cos\theta_1>0$ since $k_0>|{\vec k}|$ due to
$q^2>0$, we have $\eta =sign(k_0\cos\theta_1-|{\vec k}|)$. We
introduced the corresponding parameter $\varepsilon $ (as it was
done in the lepton--hadron scattering processes (see for example \cite{AR77}) 
\be
\varepsilon^{-1}=\frac{q^2}{2E^2_1sin^2\theta_1}-1.
\label{eq:eqepsi} 
\ee

Let us do the following remarks on the expression (\ref{eq:eqd3s}) for the
differential cross section:
\begin{itemize}
\item Its form is not related to a particular
reaction mechanism. It is a consequence of a one-photon-exchange
mechanism and of the P-invariance of the hadron electromagnetic
interaction.
\item  Its general nature is due to the fact that
its derivation requires only the hadron electromagnetic current
conservation and the fact that the photon has spin one.
\item As $q^2>>4m_e^2$ ($m_e$ is the electron mass), the components of the leptonic tensor are calculated in the limit
of zero electron mass.
\item  It holds for the case of the longitudinally
polarized electron and arbitrary polarization of the antiproton beam
and proton target.
\item The dependence of the differential cross section on two
kinematical variables, which do not characterize the mechanism of
the $\bar p+p\to \pi^0+\gamma^*$ reaction, $\phi $ (the angle
between hadron reaction plane and lepton pair production plane) and
parameter $\varepsilon $ is explicitly singled out.
\item All information about the $\bar p+p\to \pi^0+\gamma^*$ reaction
mechanism is contained in the components of the hadronic tensor
$H_{ij}$.
\end{itemize}
\subsection{The matrix element}

The electromagnetic structure of hadrons, as probed by elastic and
inelastic electron scattering, can be characterized by a set of
structure functions. Each of these structure functions is determined
by different combinations of the longitudinal and transverse
components of the electromagnetic current $J_{\mu}$, thus providing
different pieces of information about the hadron structure and the
possible mechanisms of the reaction under consideration. The formalism of the structure functions is  especially convenient for the
investigation of polarization phenomena. Similar formalism can be
applied to the annihilation reactions for the time-like region of
the virtual photon.

Taking into account the conservation of the leptonic $j_{\mu}$ and
hadronic $J_{\mu}$ electromagnetic currents, the matrix element of
the  $\bar p+p\to e^++e^-+\pi^0$ reaction can be written as
\be
\label{eq:eq14}
{\cal M}=ee_{\mu}J_{\mu}=e{\vec \epsilon}\cdot {\vec J}, \ \
e_{\mu}=\frac{e}{q^2}j_{\mu}, \ \ {\vec \epsilon}=\frac{{\vec
e}\cdot {\vec k}}{k_0^2}{\vec k}-{\vec e}.
\ee
The structure of this matrix element has to be the same as for
the $e^-+p\to e^-+p+\pi $ reaction (electroproduction of pions on
nucleon) because it is the crossed channel of the reaction (\ref{eq:eq1}).

Let us use the approach, developed for the investigation of the
polarization phenomena in the reaction of the electroproduction of
pions on nucleon \cite{AR77}, for the case of the reaction (\ref{eq:eq1}). 
 
Let us introduce, for convenience and to simplify the following calculations of the polarization
observables, in the $\bar p p$ CMS system, the orthogonal system of basic unit vectors ${\vec
q}$, ${\vec m}$ and ${\vec n}$  which are built from the momenta of
the particles participating in the reaction
\begin{equation}\label{15}
{\vec q}=\frac{{\vec k}}{|{\vec k}|}, \ \ {\vec n}=\frac{{\vec
k}\times {\vec p}}{|{\vec k}\times {\vec p}|}, \ \ {\vec m}={\vec
n}\times {\vec q}.
\end{equation}
The unit vectors ${\vec q}$ and ${\vec m}$ define the $xz$- reaction plane  for the process $\bar p+p\to
\pi+\gamma^*$ and are directed along $z$ and $x$
axes, respectively. The unit vector ${\vec n}$ is perpendicular to
the reaction plane and it is directed along the $y$ axis.

We get for the matrix element the following expression after converting from four- to two-component spinors:
\begin{equation}\label{16}
{\cal M}=e\varphi_{\pi}^+\varphi_2^+F\varphi_1,
\end{equation}
where $\varphi_1 (\varphi_2)$ is the proton (antiproton) spinor,
respectively and $F$ is the reaction amplitude which can be chosen
in different but equivalent forms.

Being the crossed channel of the pion
electroproduction on nucleon  $e+N\to e+N+\pi $, the amplitude $F$, which describes the
dynamics of the $\bar p+p\to \gamma^*+\pi^0$ reaction, is determined
in the general case by six independent amplitudes for the P-conserving
hadronic interaction. In the analysis of the polarization phenomena,
it is convenient to use the following general form for the amplitude
$F$:
\be\label{17}
F={\vec \epsilon}\cdot {\vec m}(f_1{\vec \sigma}\cdot {\vec
m}+f_2{\vec \sigma}\cdot {\vec q})+ {\vec \epsilon}\cdot {\vec
n}(if_3+f_4{\vec \sigma}\cdot {\vec n})+ {\vec \epsilon}\cdot {\vec
q}(f_5{\vec \sigma}\cdot {\vec m}+f_6{\vec \sigma}\cdot {\vec q}),
\ee 
where $f_i$, $i=1-6$ are
the scalar amplitudes in the orthogonal basis which completely
determine the reaction dynamics. These amplitudes contain
information about the dynamics of the considered reaction and they
depend on three independent kinematical variables, for example
$s=(p_1+p_2)^2$, $q^2$ and $t=(p_1-q)^2$.
\subsection{Hadronic tensor}

Let us consider the general properties of the hadronic tensor. Taking
into account that the hadronic and leptonic tensors satisfy
the following conditions: $H_{\mu\nu}q_{\mu}=H_{\mu\nu}q_{\nu}=0$,
$L_{\mu\nu}q_{\mu}=L_{\mu\nu}q_{\nu}=0$ (as a consequence of the
hadronic and leptonic currents conservation) one can show that all
polarization observables in the reaction (\ref{eq:eq1})  are
determined by the space components of the hadronic tensor only.

The hadronic tensor $H_{ij}$ $(i, j=x, y, z)$ can be represented as a
sum of terms which can be classified according to the polarization states of the
proton-antiproton system, in the following way
\be
H_{ij}=H_{ij}(0)+H_{ij}(\xi_1)+H_{ij}(\xi_2)+H_{ij}(\xi_1,\xi_2),
\label{eq:14}
\ee
where the term $H_{ij}(0)$ corresponds to the case of the
unpolarized proton-antiproton system, the term $H_{ij}(\xi_1)
(H_{ij}(\xi_2))$ corresponds to the case of polarized proton
(antiproton) and the term $H_{ij}(\xi_1,\xi_2)$ corresponds to the
case when both initial particles are polarized.

The general structure of the first term of the hadronic tensor in Eq. (\ref{eq:14}), which
corresponds to unpolarized antiproton and proton, has the following
form 
\be 
H_{ij}(0)=\alpha_1q_iq_j+\alpha_2n_in_j+\alpha_3m_im_j+
\alpha_4(q_im_j+q_jm_i)+i\alpha_5(q_im_j-q_jm_i). 
\label{eq:Hij0}
\ee 
The real structure functions $\alpha_i$, $i=$1-5, depend on
three variables, for example $s,$ $q^2$ and $\cos\theta $ where $\theta $ is the
angle between the momenta of the antiproton and the virtual photon.
Note that the structure function $\alpha_5 $, for the crossed
(scattering) channel $e+N \to e+N+\pi $, is determined by the strong
interaction effects of the final-state hadrons and vanishes for the
pole diagrams contribution (the Born approximation) in all the
kinematical range. In our case this structure function is non zero
even in the Born approximation due to the fact that hadron form
factors are complex here. This structure function gives contribution
to the cross section only in the case of polarized leptons, where
the lepton tensor contains an antisymmetric part. The structure
functions are related to the reaction scalar amplitudes $f_i$
\cite{Gakh:2010qp}. The expressions of these structure functions in
terms of the scalar amplitudes are given in Appendix \ref{AppA}.

Let us consider the case when only one hadron is polarized in the
initial state. Then the general structure of the polarized hadronic tensor can be written as
\ba
H_{ij}(\xi_1 )&=&{\vec \xi_1}\cdot {\vec n}\Bigl
(\beta_1q_iq_j+\beta_2m_im_j+
\beta_3n_in_j+\beta_4\{q,m\}_{ij}+i\beta_5[q,m]_{ij}\Bigr )+
\nn\\
&&
+{\vec \xi_1}\cdot {\vec q}\Bigl (\beta_6\{q,n\}_{ij}+\beta_7\{m,n\}_{ij}+
i\beta_8[q,n]_{ij}+i\beta_9[m,n]_{ij}\Bigr )+ \nn\\
&&
+{\vec \xi_1}\cdot {\vec m}\Bigl (\beta_{10}\{q,n\}_{ij}+\beta_{11}\{m,n\}_{ij}+
i\beta_{12}[q,n]_{ij}+i\beta_{13}[m,n]_{ij}\Bigr ),
\label{eq:Hijx}
\ea
where ${\vec
\xi_1}$ is the unit vector of the proton polarization and
$\{a,b\}_{ij}=a_ib_j+a_jb_i$, $[a,b]_{ij}=a_ib_j-a_jb_i. $

We denote as $\bar\beta_i$ the structure functions that determine
the hadronic tensor $H_{ij}(\xi_2)$, where ${\vec \xi}_2$ is the
antiproton polarization unit vector, corresponding to the case when
the antiproton is polarized. The structure of this hadronic tensor
is similar to the case of polarized proton and is given by Eq.
(\ref{eq:Hijx}), under replacement $\beta_i\to \bar\beta_i$, and $\xi_1\to \xi_2$.

Therefore, the dependence of the polarization observables on the
nucleon (or antinucleon) polarization is determined by 13 structure
functions. The expressions of these structure functions in terms of
the reaction scalar amplitudes are also given in Appendix \ref{AppA}.

Let us consider the case when both initial particles are polarized.
The general structure of the hadronic tensor which describes the
correlation of the nucleon and antinucleon polarizations can be
written as
\ba
H_{ij}(\xi_1 ,\xi_2 )&=&{\vec \xi_1}\cdot {\vec m}{\vec \xi_2}\cdot
{\vec m} \Bigl (\gamma_1q_iq_j+\gamma_2m_im_j+
\gamma_3n_in_j+\gamma_4\{q,m\}_{ij}+i\gamma_5[q,m]_{ij}\Bigr )+
\nn\\
&&{\vec \xi_1}\cdot {\vec m}{\vec \xi_2}\cdot {\vec q}
\Bigl (\gamma_6q_iq_j+\gamma_7m_im_j+
\gamma_8n_in_j+\gamma_9\{q,m\}_{ij}+i\gamma_{10}[q,m]_{ij}\Bigr )+
\nn\\
&&{\vec \xi_1}\cdot {\vec q}{\vec \xi_2}\cdot {\vec m}
\Bigl (\gamma_{11}q_iq_j+\gamma_{12}m_im_j+
\gamma_{13}n_in_j+\gamma_{14}\{q,m\}_{ij}+i\gamma_{15}[q,m]_{ij}\Bigr
)+\nn\\
&&
{\vec \xi_1}\cdot {\vec n}{\vec \xi_2}\cdot {\vec n}
\Bigl (\gamma_{16}q_iq_j+\gamma_{17}m_im_j+
\gamma_{18}n_in_j+\gamma_{19}\{q,m\}_{ij}+i\gamma_{20}[q,m]_{ij}\Bigr
)+\nn\\
&&
{\vec \xi_1}\cdot {\vec q}{\vec \xi_2}\cdot {\vec q}
\Bigl (\gamma_{21}q_iq_j+\gamma_{22}m_im_j+
\gamma_{23}n_in_j+\gamma_{24}\{q,m\}_{ij}+i\gamma_{25}[q,m]_{ij}\Bigr
)+\nn\\
&&
{\vec \xi_1}\cdot {\vec m}{\vec \xi_2}\cdot {\vec n}
\Bigl (\gamma_{26}\{q,n\}_{ij}+\gamma_{27}\{m,n\}_{ij}+
i\gamma_{28}[q,n]_{ij}+i\gamma_{29}[m,n]_{ij}\Bigr )+
\nn\\
&&
{\vec \xi_1}\cdot {\vec n}{\vec \xi_2}\cdot {\vec m}
\Bigl (\gamma_{30}\{q,n\}_{ij}+\gamma_{31}\{m,n\}_{ij}+
i\gamma_{32}[q,n]_{ij}+i\gamma_{33}[m,n]_{ij}\Bigr )+\nn\\
&&
{\vec \xi_1}\cdot {\vec n}{\vec \xi_2}\cdot {\vec q}
\Bigl (\gamma_{34}\{q,n\}_{ij}+\gamma_{35}\{m,n\}_{ij}+
i\gamma_{36}[q,n]_{ij}+i\gamma_{37}[m,n]_{ij}\Bigr )+\nn\\
&&
{\vec \xi_1}\cdot {\vec q}{\vec \xi_2}\cdot {\vec n}
\Bigl (\gamma_{38}\{q,n\}_{ij}+\gamma_{39}\{m,n\}_{ij}+
i\gamma_{40}[q,n]_{ij}+i\gamma_{41}[m,n]_{ij}\Bigr ),
\label{eq:Hijxi}
\ea
where
${\vec \xi_1}({\vec \xi_2})$ is the nucleon (antinucleon)
polarization unit vector. We see that the polarization observables,
which are due to the correlation between the nucleon and antinucleon
polarizations are determined by 41 structure functions $\gamma_i
(i=1-41).$ Their expressions in terms of the reaction scalar
amplitudes $f_i (i=1-6)$ are given in Appendix \ref{AppB}.

Let us stress that the results listed above have a general nature
and are not related to a particular reaction mechanism. They are
valid for the one-photon-exchange mechanism assuming P-invariance of
the hadron electromagnetic interaction. The general nature of these results follows from the fact that their derivation requires only the
hadron electromagnetic current conservation and the spin one nature of the virtual photon.
\section{Cross section and polarization observables}

Let us calculate the five-fold differential cross section and various
polarization observables for different types of experiments.
\subsection{Unpolarized cross section}

Let us calculate the differential cross section of the reaction
(\ref{eq:eq1}) for the case when all particles are unpolarized. It
is necessary to calculate the components of the hadronic tensor
$H_{ij}(0)$ which general form is given in Eq. (\ref{eq:Hij0}). One
has 
\ba 
H_{xx}(0)\pm H_{yy}(0)&=&\alpha_3\pm \alpha_2, \
H_{zz}(0)=\alpha_1, \nn H_{xz}(0)+H_{zx}(0)=2\alpha_4, \\
H_{xz}(0)-H_{zx}(0)&=&-2i\alpha_5. 
\label{eq:Hxz} 
\ea 
The remaining
components vanish. 

The general form of the differential cross section for the reaction
$\bar p+p\to e^++e^-+\pi^0$ in the case of unpolarized particles for
the experimental conditions where the pion and the electron are detected
in coincidence, can be written in terms of four independent
contributions (the average over the spins of the initial particles
is taken into account)
\ba
\frac{d^3\sigma_{un}}{dE_{\pi}d\Omega_{\pi}d\Omega_e} &=&N\Sigma,~
\Sigma =\sigma_{T}+\varepsilon \cos2\phi \sigma_{P}+\varepsilon
\sigma_{L}+ \sqrt{2\varepsilon (1-\varepsilon )}\cos\phi \sigma_{I},
\nn\\
\sigma_{T}&=&|f_1|^2+|f_2|^2+|f_3|^2+|f_4|^2, \ \sigma_{L}=2\frac{q^2}{k_0^2}
(|f_5|^2+|f_6|^2), \nn\\
\sigma_{P}&=&|f_3|^2+|f_4|^2-|f_1|^2-|f_2|^2,~
\sigma_{I}=-2\frac{\sqrt{q^2}}{k_0}Re(f_1f_5^*+f_2f_6^*)\eta .
\label{eq:d3sun}
\ea
Note that the scalar amplitudes $f_{i=1-4}$ are
determined by the transverse components of the 
hadron electromagnetic current whereas $f_5$, $f_6$ are related to the longitudinal part. The subscripts $L$ and $T$ indicate that the
corresponding contributions are determined by the longitudinal and
transverse components of the electromagnetic current, respectively.
The contribution $\sigma_P (\sigma_I)$ is determined by the
interference of the transverse-transverse (transverse-longitudinal)
components of this current. For
the annihilation channel, Eq. (\ref{eq:d3sun}) 
is similar to the corresponding expression in the scattering
channel for the reactions of the following type $e+A\to e+h+B$ (for example, $e+N\to
e+\pi+N$ or $e+d\to e+n+p$) for an experimental setup where the
scattered electron and the produced hadron $h$ are detected in
coincidence \cite{AR77}.

Since the differential cross section depends on the azimuthal angle
$\phi $ one can define the following asymmetry (the so-called
left-right asymmetry in the case of the scattering channel)
\be
A_{LT}=\frac{d\sigma(\phi=0^{\circ})-d\sigma(\phi=180^{\circ})}
{d\sigma(\phi=0^{\circ})+d\sigma(\phi=180^{\circ})}
=\frac{\sqrt{2\epsilon (1-\epsilon )}\sigma_I }{\sigma_T+\epsilon
(\sigma_L+\sigma_P)}.
\label{eq:ALT}
\ee
This asymmetry is determined by the interference of the reaction
amplitudes which characterize the emission of the virtual photon
with nonzero longitudinal and transverse polarizations (note the
subscript $LT$). In the annihilation channel, we show below that this
asymmetry is proportional to $\sin\theta $.
\subsection{Longitudinally polarized electron}

The differential cross section for the case when the produced electron
is longitudinally polarized can be written as
\be
\frac{d^3\sigma}{dE_{\pi}d\Omega_{\pi}d\Omega_e}
=\frac{1}{2}\frac{d^3\sigma_{un}}{dE_{\pi}d\Omega_{\pi}d\Omega_e}
(1+\lambda A'_{LT}),
\label{eq:eqpe}
\ee
where the asymmetry $A'_{LT}$ is determined as follows
\ba
A'_{LT}&=&\frac{d\sigma(\lambda=1)-d\sigma(\lambda=-1)}
{d\sigma(\lambda=1)+d\sigma(\lambda=-1)} =\sqrt{2\varepsilon
(1+\varepsilon )}\sigma'_I \Sigma^{-1},\nn\\
\sigma'_I&=&2\frac{\sqrt{q^2}}{k_0}Im(f_1f_5^*+f_2f_6^*).
\label{eq:ALT1}
\ea
Note that, due to the specific $\phi $-dependence, this asymmetry
has to be measured in noncoplanar geometry (out-of-plane
kinematics). One can show that this asymmetry is also proportional
to $\sin\theta $.
\subsection{Longitudinally polarized electron and polarized proton target or polarized antiproton beam}
Let us consider the production of the longitudinally polarized electron in
the annihilation reaction (\ref{eq:eq1}) where either the proton target or the antiproton beam are polarized. 

The part of the differential cross section which depends on the
polarization of the proton target can be written as:
\be
\frac{d^3\sigma}{dE_{\pi}d\Omega_{\pi}d\Omega_e}
=\frac{1}{2}\frac{d^3\sigma_{un}}{dE_{\pi}d\Omega_{\pi}d\Omega_e}
\bigl [1+(A_x^p+\lambda T_x^p)\xi_{1x}+(A_y^p+\lambda
T_y^p)\xi_{1y}+(A_z^p+\lambda T_z^p)\xi_{1z}\bigr ],
\label{eq:eq27}
\ee
where $A_i^p$, $i=x,y,z,$ are the analyzing powers (or the
asymmetries) due to the polarization of the proton target, and
$T_i^p$, $i=x,y,z,$ are the coefficients of the polarization transfer
from the polarized proton to the produced electron.

One can explicitly single out the dependence of the polarization
observables $A_i^p$ and $T_i^p$ on the azimuthal angle $\phi $ and
parameter $\varepsilon $ and write the expressions for these
observables as functions of the contributions of the longitudinal
(L) and transverse (T) components of the hadronic current in the
$\bar p+p\to \gamma^*+\pi^{\circ}$ reaction
\ba
\Sigma A_x^p&=&\sin\phi \bigl [\sqrt{2\varepsilon (1-\varepsilon
)}A_x^{p(LT)}+\varepsilon \cos\phi A_x^{p(TT)}\bigr ], \nn\\
\Sigma A_y^p&=&A_y^{p(TT)}+\varepsilon A_y^{p(LL)}+
\sqrt{2\varepsilon (1-\varepsilon )}\cos\phi  A_y^{p(LT)}+
\varepsilon \cos2\phi \bar A_y^{p(TT)}, \nn\\
\Sigma A_z^p&=&\sin\phi \bigl [\sqrt{2\varepsilon (1-\varepsilon
)}A_z^{p(LT)}+\varepsilon \cos\phi A_z^{p(TT)}\bigr ],\nn\\
\Sigma T_x^p&=&\sqrt{2\varepsilon (1+\varepsilon
)}\cos\phi T_x^{p(LT)}+\sqrt{1-\varepsilon^2}T_x^{p(TT)}, \nn\\
\Sigma T_y^p&=&\sqrt{2\varepsilon (1+\varepsilon
)}\sin\phi T_y^{p(LT)}, \nn\\
\Sigma T_z^p&=&\sqrt{2\varepsilon (1+\varepsilon
)}\cos\phi T_z^{p(LT)}+\sqrt{1-\varepsilon^2}T_z^{p(TT)}, 
\label{eq:eqAT}
\ea
where the individual contributions to the above polarization observables
in terms of the structure functions $\beta_i$ are given by
\ba
A_x^{p(TT)}&=&-4\beta_{11}, \ A_z^{p(TT)}=-4\beta_{7},
\ A_x^{p(LT)}=-2\beta_{10}\eta \frac{\sqrt{q^2}}{k_0}, \ \
A_z^{p(LT)}=-2\beta_{6}\eta \frac{\sqrt{q^2}}{k_0}, \nn\\
A_y^{p(TT)}&=&\beta_{2}+\beta_{3}, \ \bar
A_y^{p(TT)}=\beta_{3}-\beta_{2}, \
A_y^{p(LL)}=2\beta_{1}\frac{q^2}{k_0^2}, \
A_y^{p(LT)}=-2\beta_{4}\eta \frac{\sqrt{q^2}}{k_0},  \nn\\
T_x^{p(TT)}&=&-2\beta_{13}\eta , \
T_x^{p(LT)}=2\beta_{12}\frac{\sqrt{q^2}}{k_0}, \
T_y^{p(LT)}=-2\beta_{5}\frac{\sqrt{q^2}}{k_0},
 \nn\\
T_z^{p(TT)}&=&-2\beta_{9}\eta , \
T_z^{p(LT)}=2\beta_{8}\frac{\sqrt{q^2}}{k_0}.
\label{eq:eqAT1}
\ea
The structure
functions $\beta_i$ describe the polarization phenomena for the case
when the proton target is polarized. Similarly, the polarization effects for
the case of polarized antiproton beam are described by the
structure functions $\bar\beta_i, i=1-13.$ Their expressions in
terms of the reaction scalar amplitudes are given in Appendix \ref{AppB}.

The part of the differential cross section which depends on the
polarization of the antiproton beam has the form given by Eq. (\ref{eq:eq27}), 
where it is necessary to do the following substitution: ${\vec
\xi}_1\to {\vec \xi}_2,$ $A_i^p\to A_i^{\bar p}$, and $T_i^p\to
T_i^{\bar p}$. The dependence of the polarization observables
$A_i^{\bar p}$ and $T_i^{\bar p}$ on the azimuthal angle $\phi $ and
parameter $\varepsilon $ is the same as in Eq. (\ref{eq:eqAT}). The expressions 
of the individual contributions $A_i^{\bar p(MN)}$ and $T_i^{\bar
p(MN)}, MN=LL, LT, TT,$ are given in Eq. (\ref{eq:eqAT1}), where it is necessary
to do the substitution $\beta_i\to\bar\beta_i$.

\subsection{Longitudinally polarized electron, polarized proton target
and antiproton beam}

The part of the differential cross section which depends on the
product of the polarizations of the proton target and antiproton
beam can be written as
\ba
\frac{d^3\sigma}{dE_{\pi}d\Omega_{\pi}d\Omega_e}
&=&\frac{1}{2}\frac{d^3\sigma_{un}}{dE_{\pi}d\Omega_{\pi}d\Omega_e}
\bigl [1+(C_{xx}+\lambda \bar
C_{xx})\xi_{1x}\xi_{2x}+(C_{yy}+\lambda \bar
C_{yy})\xi_{1y}\xi_{2y}+ \nn\\
&&(C_{zz}+\lambda \bar C_{zz})\xi_{1z}\xi_{2z}+
(C_{xy}+\lambda \bar C_{xy})\xi_{1x}\xi_{2y}+
(C_{xz}+\lambda \bar
C_{xz})\xi_{1x}\xi_{2z}+\nn\\
&&(C_{yx}+\lambda \bar
C_{yx})\xi_{1y}\xi_{2x}+  \nn\\
&&(C_{yz}+\lambda \bar C_{yz})\xi_{1y}\xi_{2z}+
(C_{zx}+\lambda \bar C_{zx})\xi_{1z}\xi_{2x}+ \nn\\
&&(C_{zy}+\lambda \bar
C_{zy})\xi_{1z}\xi_{2y}\bigr ],
\label{eq:eqcxx1}
\ea
where $C_{ij}$, $i,j=x, y, z,$ are
the double spin correlation coefficients and $\bar C_{ij}$ $i,j=x, y, z,$ are the
triple spin coefficients describing the dependence of the longitudinal
polarization of the electron on the polarization state of the
polarized initial particles.

One can explicitly single out the dependence of the polarization
observables $C_{ij}$ and $\bar C_{ij}$ on the azimuthal angle $\phi
$ and parameter $\varepsilon $ and write the expressions for these
observables as functions of the contributions of the longitudinal
(L) and transverse (T) components of the hadronic current in the
$\bar p+p\to \gamma^*+\pi^{\circ}$ reaction as:
\ba
\Sigma C_{xx}&=&C_{xx}^{(TT)}+\varepsilon C_{xx}^{(LL)}+
\sqrt{2\varepsilon (1-\varepsilon )}\cos\phi  C_{xx}^{(LT)}+
\varepsilon \cos2\phi \bar C_{xx}^{(TT)},  \nn\\
\Sigma C_{yy}&=&C_{yy}^{(TT)}+\varepsilon C_{yy}^{(LL)}+
\sqrt{2\varepsilon (1-\varepsilon )}\cos\phi  C_{yy}^{(LT)}+
\varepsilon \cos2\phi \bar C_{yy}^{(TT)},  \nn\\
\Sigma C_{zz}&=&C_{zz}^{(TT)}+\varepsilon C_{zz}^{(LL)}+
\sqrt{2\varepsilon (1-\varepsilon )}\cos\phi  C_{zz}^{(LT)}+
\varepsilon \cos2\phi \bar C_{zz}^{(TT)},  \nn\\
\Sigma C_{xz}&=&C_{xz}^{(TT)}+\varepsilon C_{xz}^{(LL)}+
\sqrt{2\varepsilon (1-\varepsilon )}\cos\phi  C_{xz}^{(LT)}+
\varepsilon \cos2\phi \bar C_{xz}^{(TT)},  \nn\\
\Sigma C_{zx}&=&C_{zx}^{(TT)}+\varepsilon C_{zx}^{(LL)}+
\sqrt{2\varepsilon (1-\varepsilon )}\cos\phi  C_{zx}^{(LT)}+
\varepsilon \cos2\phi \bar C_{zx}^{(TT)},  \nn\\
\Sigma C_{xy}&=&\sin\phi \bigl [\varepsilon \cos\phi C_{xy}^{(TT)}+
\sqrt{2\varepsilon (1-\varepsilon )}C_{xy}^{(LT)}\bigl ], \nn\\
\Sigma
C_{yx}&=&\sin\phi \bigl [\varepsilon \cos\phi C_{yx}^{(TT)}+
\sqrt{2\varepsilon (1-\varepsilon )}C_{yx}^{(LT)}\bigl ],
\nn\\
\Sigma C_{yz}&=&\sin\phi \bigl
[\varepsilon \cos\phi C_{yz}^{(TT)}+ \sqrt{2\varepsilon
(1-\varepsilon )}C_{yz}^{(LT)}\bigl ], \nn\\
\Sigma C_{zy}&=&\sin\phi \bigl
[\varepsilon \cos\phi C_{zy}^{(TT)}+ \sqrt{2\varepsilon
(1-\varepsilon )}C_{zy}^{(LT)}\bigl ], \nn\\
\Sigma \bar C_{xx}&=&\sqrt{2\varepsilon (1+\varepsilon )}
\sin\phi  \bar C_{xx}^{(LT)},
\Sigma \bar C_{yy}=\sqrt{2\varepsilon
(1+\varepsilon )} \sin\phi  \bar C_{yy}^{(LT)},
\nn\\ \Sigma \bar
C_{zz}&=&\sqrt{2\varepsilon (1+\varepsilon )} \sin\phi  \bar
C_{zz}^{(LT)},
\Sigma \bar C_{xy}=\sqrt{1-\varepsilon^2}\bar C_{xy}^{(TT)}+
\sqrt{2\varepsilon (1+\varepsilon )}\cos\phi \bar C_{xy}^{(LT)},
 \nn\\
\Sigma \bar C_{yx}&=&\sqrt{1-\varepsilon^2}\bar C_{yx}^{(TT)}+
\sqrt{2\varepsilon (1+\varepsilon )}\cos\phi \bar C_{yx}^{(LT)},  \nn\\
\Sigma \bar C_{yz}&=&\sqrt{1-\varepsilon^2}\bar C_{yz}^{(TT)}+
\sqrt{2\varepsilon (1+\varepsilon )}\cos\phi \bar C_{yz}^{(LT)},\nn\\
\Sigma \bar C_{zy}&=&\sqrt{1-\varepsilon^2}\bar C_{zy}^{(TT)}+
\sqrt{2\varepsilon (1+\varepsilon )}\cos\phi \bar C_{zy}^{(LT)},  \nn\\
\Sigma \bar C_{xz}&=&\sqrt{2\varepsilon (1+\varepsilon )}
\sin\phi  \bar C_{xz}^{(LT)}, \ \Sigma \bar C_{zx}=\sqrt{2\varepsilon
(1+\varepsilon )} \sin\phi  \bar C_{zx}^{(LT)},
\label{eq:eqsc}
\ea
where the
individual contributions to the above polarization observables in
terms of the structure functions $\gamma_i$ are given by
\ba
C_{xx}^{(TT)}&=&2(\gamma_2+\gamma_3), \ \bar
C_{xx}^{(TT)}=2(\gamma_3-\gamma_2),
C_{xx}^{(LL)}=4\gamma_1\frac{q^2}{k_0^2}, \
C_{xx}^{(LT)}=-4\gamma_4\eta \frac{\sqrt{q^2}}{k_0}, \nn\\
C_{yy}^{(TT)}&=&2(\gamma_{17}+\gamma_{18}), \ \bar
C_{yy}^{(TT)}=2(\gamma_{18}-\gamma_{17}),
C_{yy}^{(LL)}=4\gamma_{16}\frac{q^2}{k_0^2}, \
C_{yy}^{(LT)}=-4\gamma_{19}\eta \frac{\sqrt{q^2}}{k_0}, \nn\\
C_{zz}^{(TT)}&=&2(\gamma_{22}+\gamma_{23}), \ \bar
C_{zz}^{(TT)}=2(\gamma_{23}-\gamma_{22}),
C_{zz}^{(LL)}=4\gamma_{21}\frac{q^2}{k_0^2}, \
C_{zz}^{(LT)}=-4\gamma_{24}\eta \frac{\sqrt{q^2}}{k_0}, \nn\\
C_{xz}^{(TT)}&=&2(\gamma_7+\gamma_8), \ \bar
C_{xz}^{(TT)}=2(\gamma_8-\gamma_7),
C_{xz}^{(LL)}=4\gamma_6\frac{q^2}{k_0^2}, \
C_{xz}^{(LT)}=-4\gamma_9\eta \frac{\sqrt{q^2}}{k_0}, \nn\\
C_{zx}^{(TT)}&=&2(\gamma_{12}+\gamma_{13}), \ \bar
C_{zx}^{(TT)}=2(\gamma_{13}-\gamma_{12}),
C_{zx}^{(LL)}=4\gamma_{11}\frac{q^2}{k_0^2}, \
C_{zx}^{(LT)}=-4\gamma_{14}\eta \frac{\sqrt{q^2}}{k_0}, \nn\\
C_{xy}^{(TT)}&=&-8\gamma_{27}, \ \ C_{xy}^{(LT)}=-4\gamma_{26}\eta
\frac{\sqrt{q^2}}{k_0}, \ C_{yx}^{(TT)}=-8\gamma_{31}, \ \
C_{yx}^{(LT)}=-4\gamma_{30}\eta \frac{\sqrt{q^2}}{k_0}, \nn\\
C_{yz}^{(TT)}&=&-8\gamma_{35}, \ \ C_{yz}^{(LT)}=-4\gamma_{34}
\frac{\sqrt{q^2}}{k_0}, \ C_{zy}^{(TT)}=-8\gamma_{39},
C_{zy}^{(LT)}=-4\gamma_{38}\eta \frac{\sqrt{q^2}}{k_0},\nn\\
\bar C_{xx}^{(LT)}&=&-4\gamma_5\frac{\sqrt{q^2}}{k_0}, \ \bar
C_{yy}^{(LT)}=-4\gamma_{20}\frac{\sqrt{q^2}}{k_0}, \ \bar
C_{zz}^{(LT)}=-4\gamma_{25}\frac{\sqrt{q^2}}{k_0},\nn\\
\bar C_{xy}^{(TT)}&=&-4\gamma_{29}\eta , \ \ \bar
C_{xy}^{(LT)}=4\gamma_{28} \frac{\sqrt{q^2}}{k_0}, \ \bar
C_{yx}^{(TT)}=-4\gamma_{33}\eta , \ \ \bar
C_{yx}^{(LT)}=4\gamma_{32} \frac{\sqrt{q^2}}{k_0}, \nn\\
\bar C_{yz}^{(TT)}&=&-4\gamma_{37}\eta , \ \ \bar
C_{yz}^{(LT)}=4\gamma_{36}
\frac{\sqrt{q^2}}{k_0}, \ \bar C_{zy}^{(TT)}=-4\gamma_{41}\eta , \nn\\
\bar C_{zy}^{(LT)}&=&4\gamma_{40} \frac{\sqrt{q^2}}{k_0}, \bar
C_{xz}^{(LT)}=-4\gamma_{10}\frac{\sqrt{q^2}}{k_0}, \ \bar
C_{zx}^{(LT)}=-4\gamma_{15}\frac{\sqrt{q^2}}{k_0}. \
\label{eq:eqCxyz} 
\ea 
Note that the expressions for the structure
functions $\gamma_i$ in terms of the reaction scalar
amplitudes $f_i$ have a general nature, not depending on the
reaction mechanism. 

At this stage, the general model-independent
analysis of the polarization observables in the reaction
(\ref{eq:eq1}) is completed. To proceed further in the calculation
of the observables, one needs a model for the reaction mechanism.
\section{Born approximation}
The numerical estimation of the differential cross section and the
polarization observables depends on the model that has been chosen
to describe the $\bar p+p\to\pi+\gamma^*$ reaction. For the present
calculation of the matrix element, let us use the approach of Ref.
\cite{Ad07}, which is based on the Compton-like Feynman amplitudes
(the Born approximation). The Feynman diagrams for this reaction are
shown in Figs. 1(a) and 1(b) for lepton pair emission from the
proton and from the antiproton, respectively. The interest of these
diagrams is that they contain the $\gamma^*pp^{(*)}$ vertex, allowing to
access nucleon form factors in the 'unphysical region' of the 
transferred momentum, below the physical threshold.

\begin{figure}
\mbox{\epsfxsize=15.cm\leavevmode \epsffile{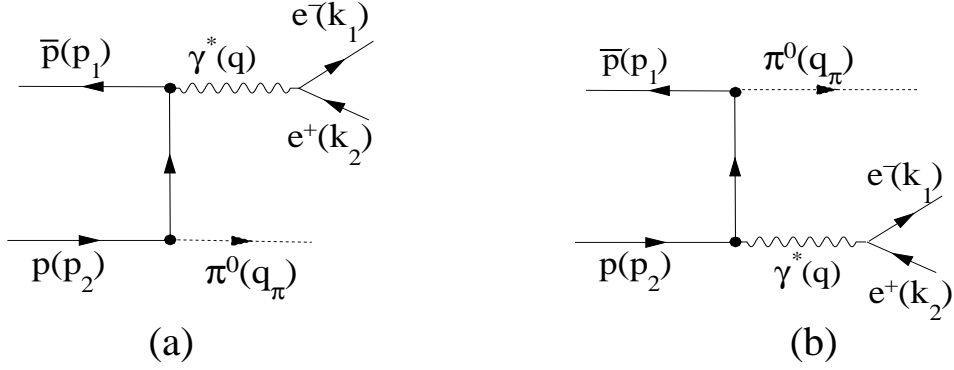}}
\caption{Feynman diagram for $\bar p+p \to e^++e^-+ \pi^0$ (a) for
lepton pair emission from the antiproton and (b) from the proton.}
\label{Fig:Fig1}
\end{figure}

As previously discussed in Refs. \cite{DDR, Ad07}, one of the
hadrons involved in the electromagnetic vertices is virtual and,
rigorously speaking, the involved form factors should be modified by
taking into account off-mass-shell effects. However, we use the
standard expression for the nucleon electromagnetic current
involving on-mass- shell nucleons, keeping in mind that the
comparison with the future data will be meaningful in the
kinematical region where the virtuality is small.

We use the following  expression for the nucleon electromagnetic
current which involves the Dirac $F_1(q^2)$ and Pauli $F_2(q^2)$
form factors
\be
<N(p_2)|\Gamma_{\mu}(q^2)|N(p_1)>=\bar
u(p_2)[F_1(q^2)\gamma_{\mu}+\frac{1}{4M} F_1(q^2)(\hat
q\gamma_{\mu}-\gamma_{\mu}\hat q)]u(p_1),
\label{eq:eqFF}
\ee
where $q$ is the four-momentum of the virtual photon. The nucleon
form factors in the kinematical region of interest for the present
work are largely unexplored complex functions.

Let us write the hadronic current corresponding to the two Feynman
diagrams of Fig. \ref{Fig:Fig1} as follows \cite{Ad07}
\be
J_{\mu}=\varphi_{\pi}^+(q_{\pi})\bar u(-p_1)O_{\mu}u(p_2).
\label{eq:Jmu}
\ee
Here $\varphi_{\pi}$ is the pion wave function. Using the Feynman
rules we can write
\be
O_{\mu}=\frac{g}{d_1}\Gamma_{\mu}(q)(\hat q-\hat p_1+M)\gamma_5+
\frac{g}{d_2}\gamma_5(\hat p_2-\hat q+M)\Gamma_{\mu}(q),
\label{eq:Omu}
\ee
where $d_i=q^2-2q\cdot p_i, i=1, 2, g$ is the coupling constant
describing the pion-nucleon vertex $\pi NN^*$, ($N^*$ is the
off-mass-shell nucleon). The possible off-mass-shell effects of this
coupling constant are neglected. Note that the hadronic current
(\ref{eq:Jmu}) is conserved, $q_{\mu}J_{\mu}=0$.

Let us write the matrix element in the proton-antiproton CMS. We
obtain the following expression for the amplitude $F$
\be
F=g_1{\vec \epsilon}\cdot {\vec k}{\vec \sigma}\cdot {\vec
k}+g_2{\vec \epsilon}\cdot {\vec k} {\vec \sigma}\cdot {\vec
p}+g_3{\vec \epsilon}\cdot {\vec p} {\vec \sigma}\cdot {\vec
k}+g_4{\vec \epsilon}\cdot {\vec p} {\vec \sigma}\cdot {\vec p}+
g_5{\vec \sigma}\cdot {\vec \epsilon}+ig_6{\vec \epsilon}\cdot
({\vec k}\times {\vec p}),
\label{eq:eqfg}
\ee
where $g_i$, $i=1-6,$ are the scalar amplitudes in a nonorthogonal
basis. The previous structure of the matrix element (\ref{eq:eqfg}) arises naturally
in the transition from four- to two-component spinors and it is also a general expression that does not depend on the details of
the reaction mechanism. In the case of the Born approximation, according to the Feynman diagrams in Fig. \ref{Fig:Fig1}, the scalar
amplitudes have the following form:
\ba
g_1^B&=&g\frac{p}{M}\left(\frac{1}{d_2}-\frac{1}{d_1}\right
)F_2(q^2),  \nn\\
g_2^B&=&\frac{g}{p}\left
(\frac{1}{d_1}-\frac{1}{d_2}\right)\left[2EF_1(q^2)+
\left(k_0+\frac{k}{M}p\cos\theta\right)F_2(q^2)\right ],  \nn\\
g_3^B&=&2g\left(\frac{1}{d_1}+\frac{1}{d_2}\right
)\left[\frac{p}{M}F_2(q^2)+ \frac{M}{p}G_M(q^2)\right ], \nn\\
g_4^B&=&-2\frac{g}{p}\left( \frac{1}{d_1}+\frac{1}{d_2}\right )
\left[(2E-k_0)F_1(q^2)+\frac{k}{M}p\cos\theta F_2(q^2)\right],  \nn\\
g_5^B&=&-2gMk\cos\theta \left(\frac{1}{d_1}+\frac{1}{d_2}\right
)G_M(q^2),
g_6^B=2g\frac{E}{p}\left(\frac{1}{d_1}+\frac{1}{d_2}\right
)G_M(q^2),
\label{eq:eqgborn}
\ea
where $G_M(q^2)=F_1(q^2)+F_2(q^2)$, $k$ $(p)$ is the
magnitude of the virtual photon (antiproton) three--momentum, and $k_0$ $
(E)$ is the energy of the virtual photon (antiproton). These
quantities are expressed in the reaction CMS as
\be
 E=\frac{\sqrt{s}}{2}, ~k_0=\frac{1}{2\sqrt{s}}(s+q^2-m_{\pi}^2),
~p^2=E ^2-M^2, ~{\vec k}^2=k_0^2-q^2,
\label{eq:eqcm}
\ee
where $m_{\pi}$ is the pion mass.

The scalar amplitudes in an orthogonal basis, $f_i$, which are used
for the analysis of the polarization effects in the reaction (\ref{eq:eq1}) can
be related to the scalar amplitudes $g_i$ through the following
relations
\ba
f_1&=&p^2\sin^2\theta g_4+g_5, \ \ f_2=p\sin\theta (kg_3+p\cos\theta
g_4), \ \ f_3=kp\sin\theta g_6,\label{eq:eqfcm}\\
f_4&=&g_5, \ \ f_5=p\sin\theta (kg_2+p\cos\theta g_4),
\ \ f_6=k^2g_1+g_5+p\cos\theta (kg_2+kg_3+p\cos\theta g_4).
\nn
\ea
\section{Numerical Results}

In a high luminosity experiment and with a detector with good efficiency
and acceptance, in particular at forward and backward angles, a precise
angular distribution of the scattered electron can be measured.

Assuming the reaction mechanism suggested in Ref. \cite{Ad07}, we
illustrate in this section the angular dependence of different
observables, for similar kinematical conditions as considered in the
previous work \cite{Gakh:2010qp}. The form factors parametrization is based on VMD model of Ref. \cite{Ia73}, and
follows Ref. \cite{TomasiGustafsson:2005kc}.

For a value of the total energy
squared $s$=5.5 GeV$^2$, three values of
the momentum transfer squared are considered: $q^2=0.5$, 2 and 4
GeV$^2$. 

\begin{figure}
\mbox{\epsfxsize=15.cm\leavevmode \epsffile{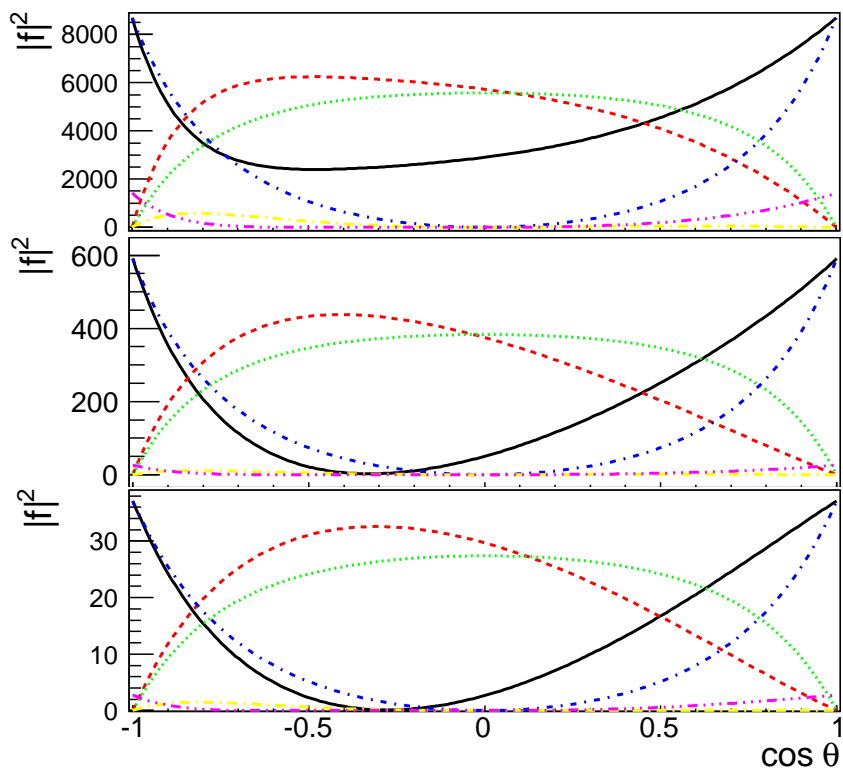}}
\caption{(Color online) Moduli squared of the amplitudes $f_i$, i=1-6, as function of $\cos\theta$ for the reaction $\bar p+p \to e^++e^-+ \pi^0$, for  $s$=5.5 GeV$^2$ and three values of
the momentum transfer squared: $q^2=0.5$ GeV$^2$,  $q^2=$2 GeV$^2$  and $q^2=4$ GeV$^2$  from top to bottom, with the following notations: $|f_1|^2$ (solid line, black), $|f_2|^2$ (dashed line, red), $|f_3|^2$ (dotted line, green), $|f_4|^2$ (dash-dotted line,blue), $|f_5|^2$ (large dash-dotted line, yellow), $|f_6|^2$ (dash-triple dotted line, magenta). }
\label{Fig:ampl}
\end{figure}

For theses values of $s$ and $q^2$, in Fig. \ref{Fig:ampl}, the moduli squared of the amplitudes are shown as functions of $\cos\theta$. The amplitudes have a characteristic angular dependence and decrease when the transferred momentum increases. They are of the same order of magnitude, except $|f_5|^2$ which is ten times smaller.  The amplitudes $|f_4|^2$, $|f_5|^2$ and $|f_6|^2$ vanish for $\cos\theta$=0.

For a simple illustration of the present results we chose to fix two more variables, $\theta_1=5^{\circ}$ and $\phi=45^{\circ}$, and show the dependence of the variables as functions of $\cos\theta$.

\begin{figure}
\mbox{\epsfxsize=15.cm\leavevmode \epsffile{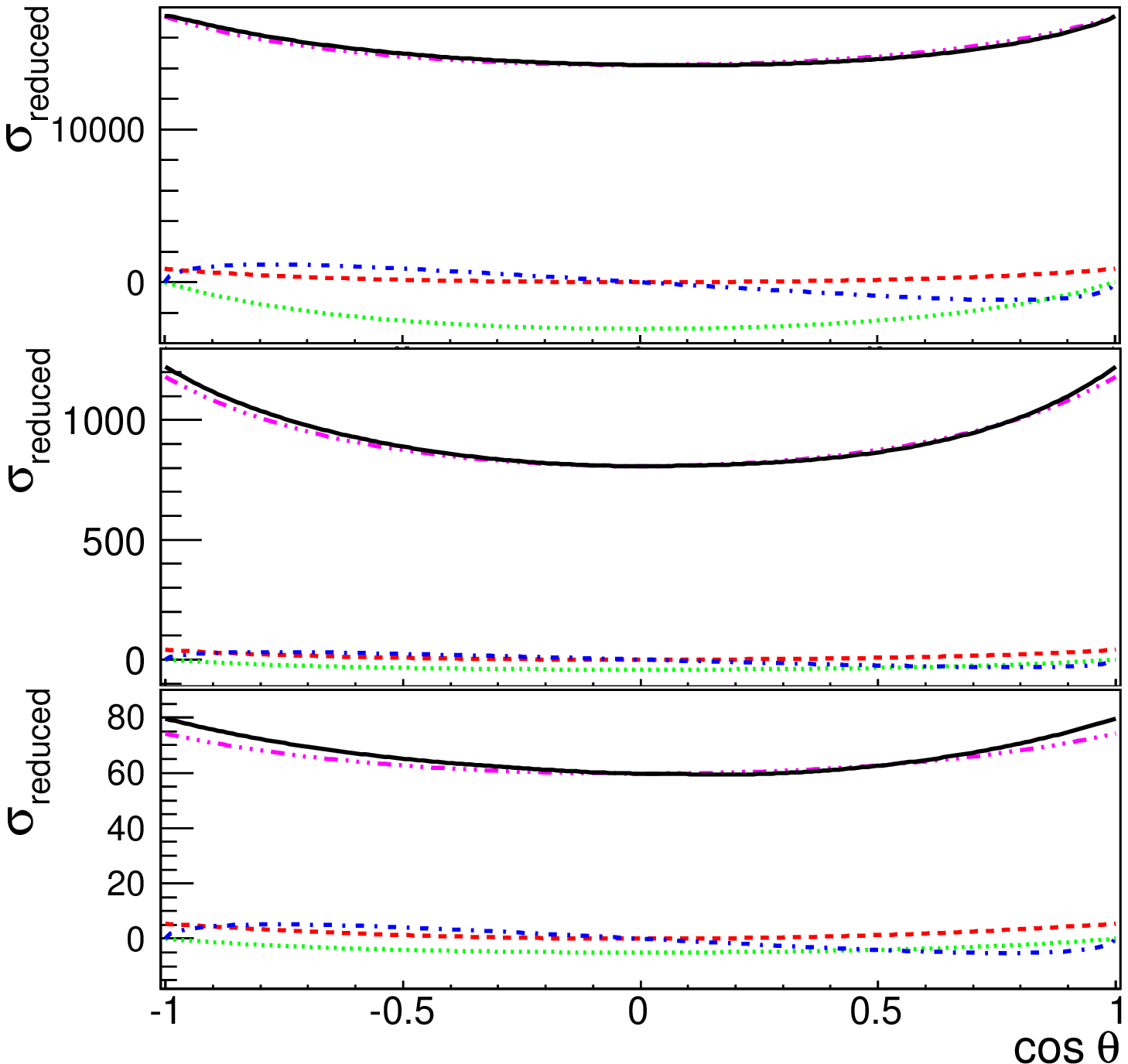}}
\caption{(Color online) Reduced cross section  $\Sigma$ ( solid line, black)  as function of $\cos\theta$ from Eq. (\protect\ref{eq:d3sun}) for the reaction $\bar p+p \to e^++e^-+ \pi^0$, at  $s$=5.5 GeV$^2$, $\theta_1=5^{\circ}$ and $\phi=45^{\circ}$ for three values of
the momentum transfer squared: $q^2=0.5$ GeV$^2$,  $q^2=$2 GeV$^2$  and $q^2=4$ GeV$^2$  from top to bottom. The individual contributions to the reduced cross section are also reported: $\sigma_T$  (large dash triple dotted line, magenta),  $\sigma_L$  (dashed line, red), $\sigma_P$  (dotted line, green), $\sigma_I$  (dash-dotted line, blue). }
\label{Fig:sigma}
\end{figure}

The reduced cross section, $\Sigma$ and its four contributions, Eq. (\protect\ref{eq:d3sun}), are shown in Fig. \ref{Fig:sigma}. As expected, the largest contribution to the cross section is given by the transversal component, $\sigma_T$, which is the incoherent sum of four amplitudes. The reduced cross section takes its maximum value for backward and forward scattering, in the considered kinematics.

\begin{figure}
\mbox{\epsfxsize=15.cm\leavevmode \epsffile{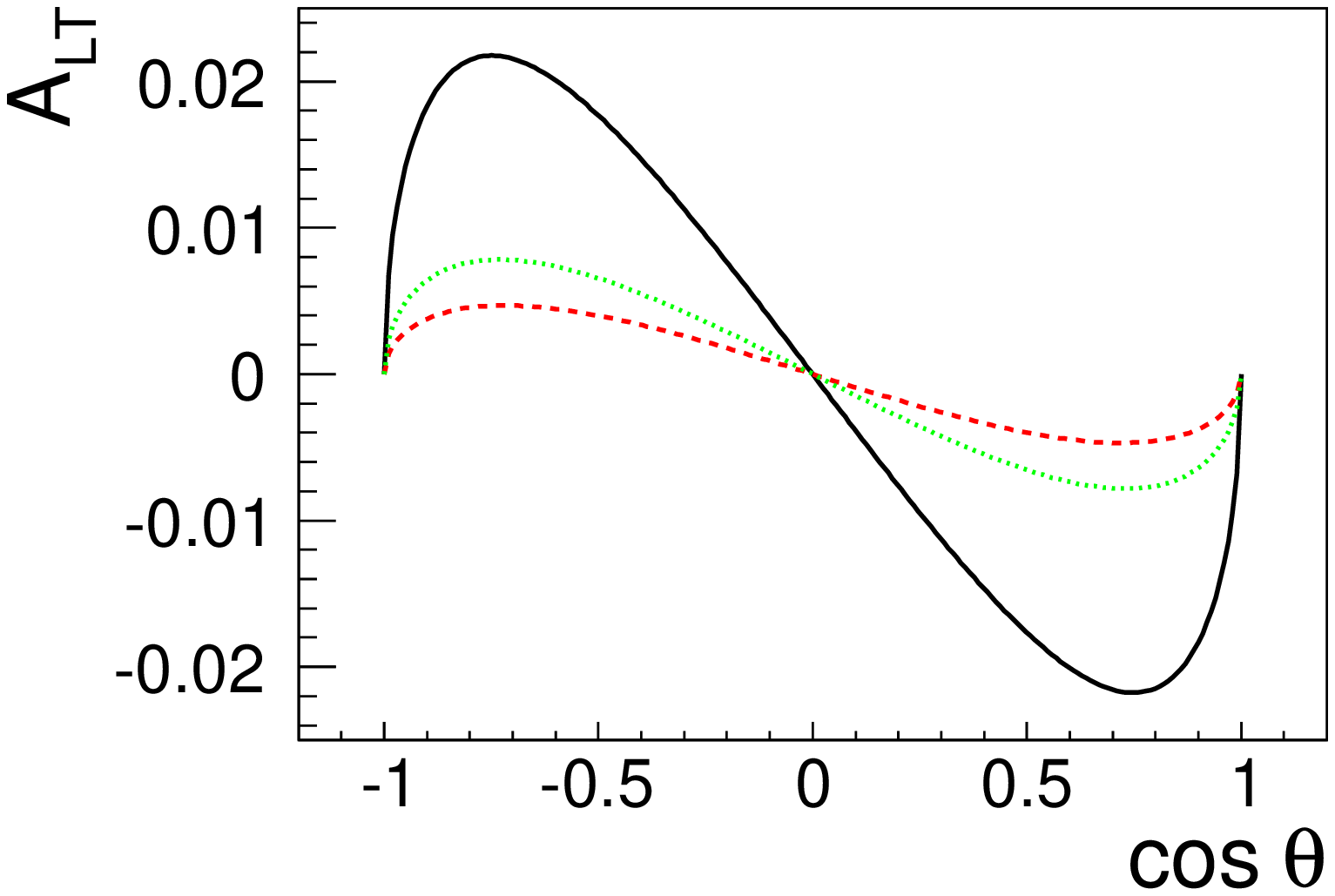}}
\caption{(Color online) Angular asymmetry (Eq. (\protect\ref{eq:ALT})) as function of $\cos\theta$ for the reaction $\bar p+p \to e^++e^-+ \pi^0$, for  $s$=5.5 GeV$^2$ and three values of
the momentum transfer squared: $q^2=0.5$ GeV$^2$(black solid line),  $q^2=$2 GeV$^2$ (dashed red line) and $q^2=4$ GeV$^2$  (green dotted line).}
\label{Fig:asym}
\end{figure}

In Fig. \ref{Fig:asym}, the angular asymmetry, Eq. (\ref{eq:ALT}),
 is illustrated as function of $\cos\theta$. It is an even function of $\cos\theta$, vanishing at forward and backward angles, and decreases when $Q^2$ increases. The single spin asymmetry, Eq. (\ref{eq:ALT1}), vanishes in Born approximation.

\begin{figure}
\mbox{\epsfxsize=15.cm\leavevmode \epsffile{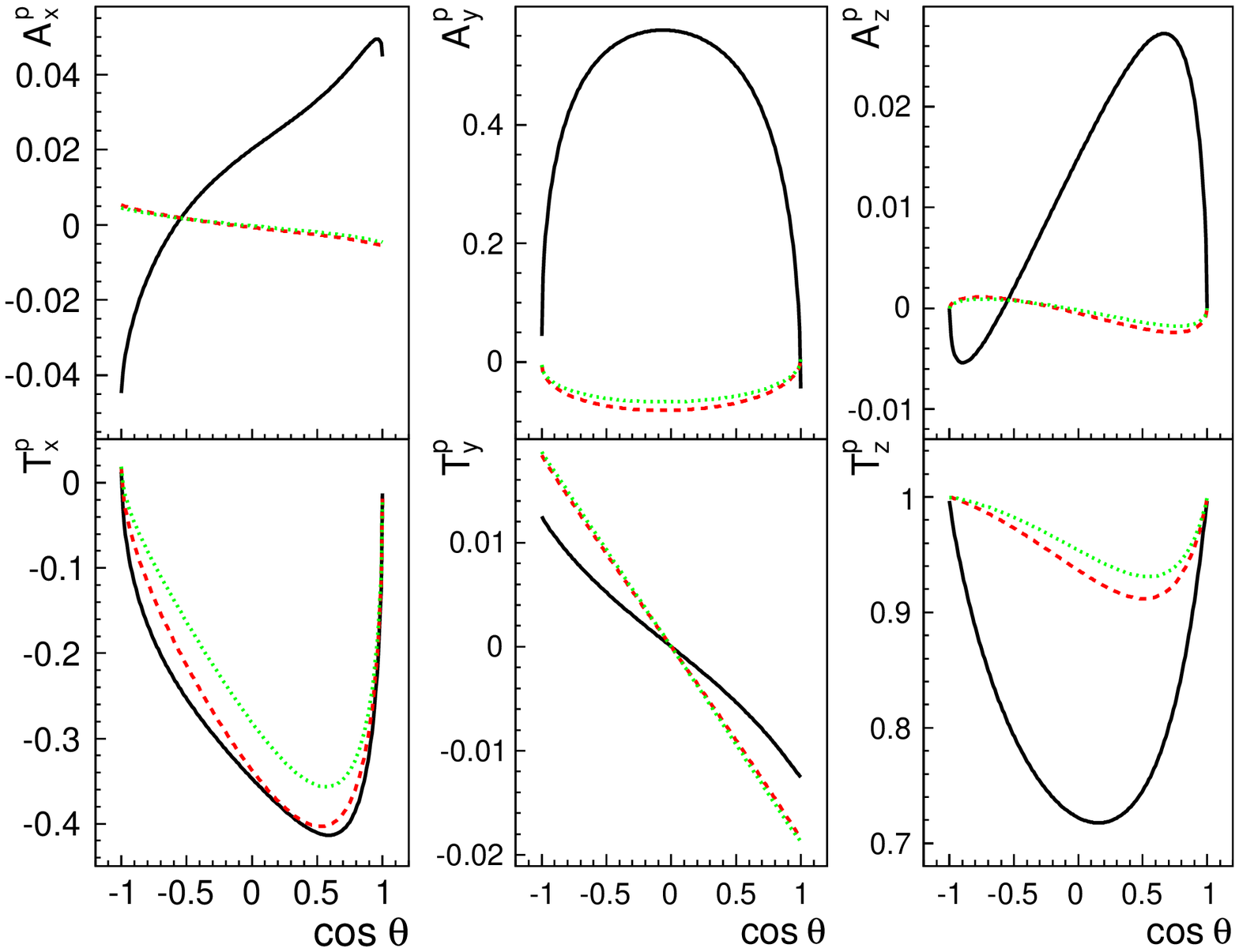}}
\caption{(Color online) Same notations as in Fig. \protect\ref{Fig:asym}, for the following polarization observables, from left to right, from top to bottom : $A_x^p$, $A_y^p$, $A_z^p$, $T_x^p$, $T_y^p$, $T_z^p$.}
\label{Fig:obsAT}
\end{figure}

The analyzing powers due to the polarization of the proton target $A_i^p, i=x,y,z,$ and
the coefficients of the polarization transfer
from the polarized proton to the produced electron  $T_i^p, i=x,y,z,$ 
(Eq. (\ref{eq:eqAT})) are illustrated in Fig. \ref{Fig:obsAT} (same notations as Fig. \ref{Fig:asym}).

\begin{figure}
\mbox{\epsfxsize=15.cm\leavevmode \epsffile{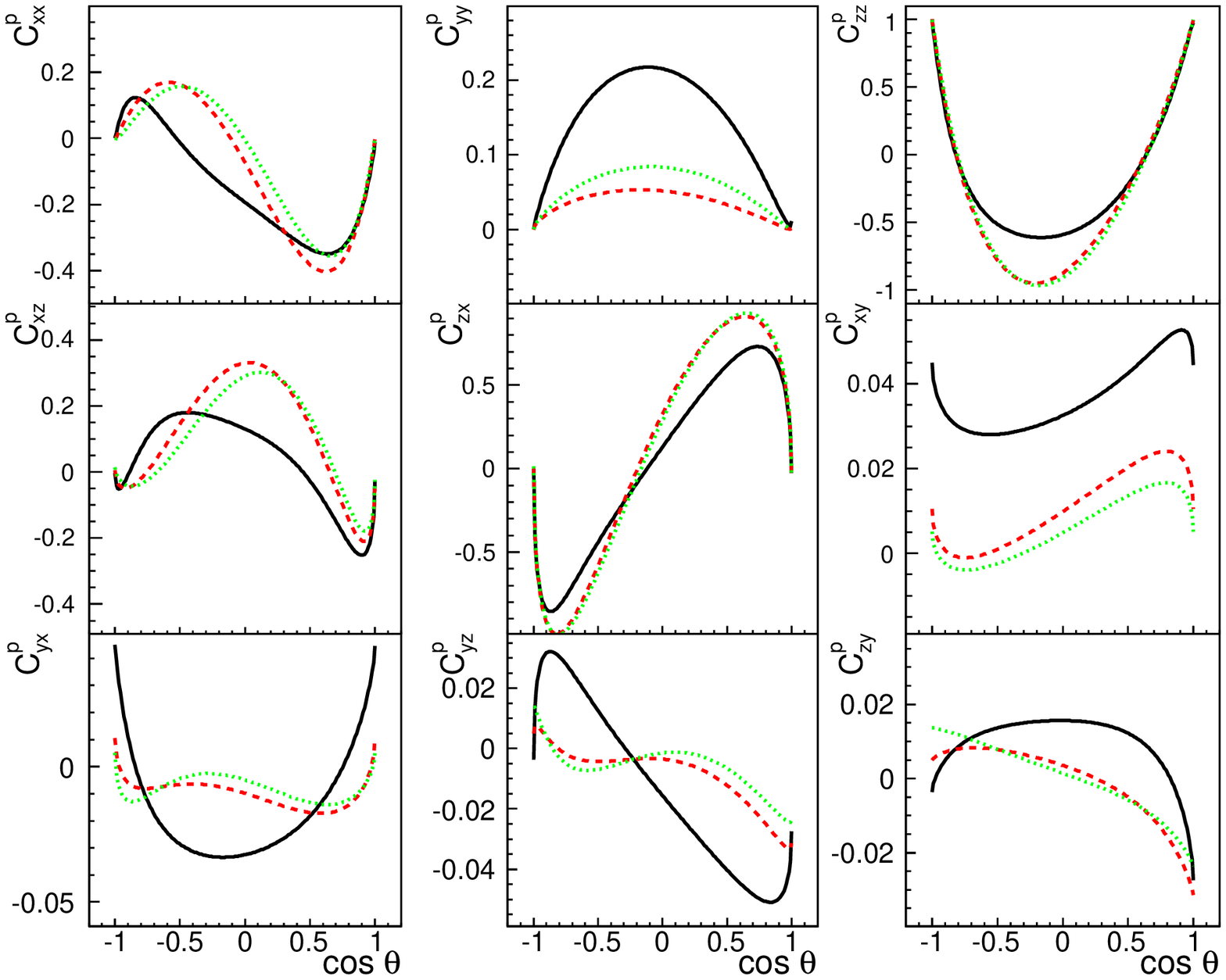}}
\caption{(Color online) Same as in Fig. \protect\ref{Fig:asym}, for the following  polarization observables  from left to right, from top to bottom : $C_{xx}^p$, $C_{yy}^p$, $C_{zz}^p$, $C_{xz}^p$, $C_{zx}^p$, $C_{xy}^p$, $C_{yx}^p$, $C_{yz}^p$, $C_{zy}^p$. }
\label{Fig:obsC}
\end{figure}

The double spin correlation coefficients $C_{ij}$ $i,j=x, y, z,$ , Eq. (\ref{eq:eqsc}), when both initial particles are polarized, are shown in Fig. \ref{Fig:obsC}.

One can see from Figs. \ref{Fig:asym}, \ref{Fig:obsAT} and \ref{Fig:obsC} that polarization observables are generally large and show a characteristic dependence on angles and momentum transfer squared. They are generally larger at smaller $q^2$ values. The comparison with future experimental data will allow to confirm or infirm the validity of the considered model.
 
\section{Conclusions}

Model independent expressions for the five-fold cross section and polarization observables have been derived for the reaction $\bar p+p \to e^++e^-+ \pi^0$, when the annihilation occurs through the one-photon exchange mechanism.

It has been shown that this reaction is completely described by a set of six amplitudes (in general complex) which are functions of three kinematical variables.

The expressions of the cross section, of the angular (out-of-plane) asymmetry and of single and double spin observables have been derived in terms of the relevant amplitudes, in a general formalism, which holds for any model of the proton    and for any reaction mechanism.

The numerical application has been done following the model of Ref. \cite{Ad07}, which is based on $t$- and $u$-  channel exchange diagrams, i.e. those diagrams which allow, in principle, to access hadron form factors in the time-like region under the physical threshold.

The results show large and measurable values of the observables, including an angular asymmetry, which is measurable with out-of-plane measurements in unpolarized particle reactions.

\section{Acknowledgments}
One of us (A.D.) acknowledges the Lebanese CNRS for financial support. This work was partly supported by  CNRS-IN2P3 (France) and by the National Academy of Sciences of Ukraine under PICS n. 5419 and by GDR PH-QCD' (France).

\appendix
\section{Helicity and scalar amplitudes}
\label{AppA}
\setcounter{equation}{0}
\def\theequation{A.\arabic{equation}}

In this Appendix the helicity amplitudes and their relation with the
reaction scalar amplitudes are given.

Let us define the helicity amplitudes as follows
\be
M_{\lambda_1 \lambda_2}^{\lambda}=e\varphi_2^+(\lambda_1 )F(\lambda
)\varphi_1(\lambda_2),
\label{eq:eqmll}
\ee
where $\lambda $ and $\lambda_1 (\lambda_2)$ are the helicities of
the virtual photon and antiproton (proton). Using the expressions
for the helicity states of the corresponding particles one obtains
the following relations between the helicity amplitudes and reaction
scalar amplitudes
\ba
h_1&=&M_{++}^0=-ie\frac{\sqrt{q^2}}{k_0}(\sin\theta f_5+\cos\theta
f_6), \ \ h_2=M_{+-}^0=-ie\frac{\sqrt{q^2}}{k_0}(-\cos\theta
f_5+\sin\theta f_6), \nn\\
h_3&=&M_{++}^+=\frac{ie}{\sqrt{2}}(\sin\theta f_1+\cos\theta f_2+f_3), \
\ h_4=M_{--}^+=\frac{ie}{\sqrt{2}}(\sin\theta f_1+\cos\theta f_2-f_3),
\nn\\
h_5&=&M_{+-}^+=\frac{ie}{\sqrt{2}}(-\cos\theta f_1+\sin\theta
f_2+f_4), \ 
\nn\\
h_6&=&M_{-+}^+=\frac{ie}{\sqrt{2}}(\cos\theta
f_1-\sin\theta f_2+f_4), 
\label{eq:hf} 
\ea 
where $\lambda =\pm , 0$
means that the virtual photon helicity is $\pm 1, 0$, respectively,
and index $\lambda_1=\pm (\lambda_2=\pm )$ means that the antiproton
(proton) helicity is $\pm 1/2 (\pm 1/2)$. The remaining helicity
amplitudes are not independent and they can be obtained from the
relation 
\be 
M_{\lambda_1 \lambda_2}^{\lambda}=(-1)^{\lambda
+\lambda_2-\lambda_1} M_{-\lambda_1 -\lambda_2}^{-\lambda} .
\label{eq:Mll} 
\ee 
We present also the inverse relation, i.e., the
expressions of the reaction scalar amplitudes in terms of the
helicity amplitudes. They are 
\ba
f_1&=&-\frac{i}{e\sqrt2}[\sin\theta (h_3+h_4)-\cos\theta (h_5-h_6)],
f_2=-\frac{i}{e\sqrt2}[\cos\theta (h_3+h_4)+\sin\theta (h_5-h_6)],\nn\\
f_3&=&-\frac{i}{e\sqrt2}(h_3-h_4), \ \
f_4=-\frac{i}{e\sqrt2}(h_5+h_6), f_5=\frac{ik_0}{e\sqrt{q^2}}(\sin\theta h_1-\cos\theta h_2),
\nn\\
f_6&=&\frac{ik_0}{e\sqrt{q^2}}(\cos\theta h_1+\sin\theta h_2).
\label{eq:eqf16}
\ea

\section{Structure functions and scalar amplitudes}
\label{AppB}
\setcounter{equation}{0}
\def\theequation{B.\arabic{equation}}

In this Appendix the explicit expressions for the structure
functions $\alpha_i$ $(i=1-5)$, $\beta_i$ $(i=1-13)$ and $\gamma_i$
$(i=1-41)$ are given in terms of the orthogonal scalar amplitudes $f_i$
$(i=1-6)$ which determine the $\bar p+p\to \pi+\gamma^*$ reaction.

The structure functions $\alpha_i$ $(i=1-5)$, describing the
hadronic tensor in the case of the unpolarized proton and
antiproton, can be written as
\ba
\alpha_1&=&2\left (|f_5|^2+|f_6|^2\right ),~ \alpha_2=2\left (|f_3|^2+|f_4|^2\right ), ~
\alpha_3=2\left (|f_1|^2+|f_2|^2\right ),\nn\\
 \alpha_4&=&2Re(f_1f_5^*+f_2f_6^*),
~\alpha_5=-2Im(f_1f_5^*+f_2f_6^*).
\label{eq:eqalpha}
\ea
The expressions for the structure functions $\beta_i (i=1-13)$,
describing the hadronic tensor in the case when only proton is
polarized, are
\ba
\beta_1&=&-2Imf_5f_6^*, ~\beta_2=-2Imf_1f_2^*,
~\beta_3=-2Imf_3f_4^*, ~\beta_4=Im(f_2f_5^*-f_1f_6^*), \nn\\
\beta_5&=&Re(f_2f_5^*-f_1f_6^*),~ \beta_6=-Im(f_3f_6^*+f_4f_5^*), ~
\beta_7=Im(f_1f_4^*+f_2f_3^*), \nn\\
\beta_8&=&-Re(f_3f_6^*+f_4f_5^*), ~ \beta_9=-Re(f_1f_4^*+f_2f_3^*), ~
\beta_{10}=Im(f_4f_6^*-f_3f_5^*), \nn\\
\beta_{11}&=&Im(f_1f_3^*-f_2f_4^*),
\beta_{12}=Re(f_4f_6^*-f_3f_5^*), ~
\beta_{13}=Re(f_2f_4^*-f_1f_3^*).
\label{eq:betai}
\ea
The expressions for the structure functions $\gamma_i$ $(i=1-41)$,
describing the hadronic tensor in the case when both the proton and
antiproton are polarized (spin correlations), are
\ba
\gamma_1&=&-\frac{1}{2}\left (|f_5|^2-|f_6|^2\right ),~
\gamma_2=-\frac{1}{2}\left (|f_1|^2-|f_2|^2\right ),~
\gamma_3=-\frac{1}{2}\left (|f_3|^2-|f_4|^2\right ),~
\nn\\
\gamma_4&=&-\frac{1}{2}Re(f_1f_5^*-f_2f_6^*),  
\gamma_5=-\frac{1}{2}Im(f_2f_6^*-f_1f_5^*), 
\gamma_6=-Ref_5f_6^*,~
\gamma_7=-Ref_1f_2^*,~ 
 \nn\\
\gamma_8&=&Ref_3f_4^*,
\gamma_9=-\frac{1}{2}Re(f_1f_6^*+f_2f_5^*), 
\gamma_{10}=\frac{1}{2}Im(f_1f_6^*+f_2f_5^*),~
\nn\\
\gamma_{11}&=&-Ref_5f_6^*,~ \gamma_{12}=-Ref_1f_2^*,~
\gamma_{13}=-Ref_3f_4^*,  
\gamma_{14}=-\frac{1}{2}Re(f_1f_6^*+f_2f_5^*),~
\nn\\
\gamma_{15}&=&\frac{1}{2}Im(f_1f_6^*+f_2f_5^*), ~
\gamma_{16}=\frac{1}{2}\left(|f_5|^2+|f_6|^2\right ),~
\gamma_{17}=\frac{1}{2}\left(|f_1|^2+|f_2|^2\right ),  
\nn\\
\gamma_{18}&=&-\frac{1}{2}\left(|f_3|^2+|f_4|^2\right ),~
\gamma_{19}=\frac{1}{2}Re(f_1f_5^*+f_2f_6^*),
~\gamma_{20}=\frac{1}{2}Im(f_5f_1^*+f_6f_2^*), 
\nn\\
\gamma_{21}&=&\frac{1}{2}\left(|f_5|^2-|f_6|^2\right ),~
\gamma_{22}=\frac{1}{2}\left(|f_1|^2-|f_2|^2\right ),~
\gamma_{23}=-\frac{1}{2}\left(|f_3|^2-|f_4|^2\right ),~
\nn\\
\gamma_{24}&=&\frac{1}{2}Re(f_1f_5^*-f_2f_6^*),  
\gamma_{25}=\frac{1}{2}Im(f_5f_1^*-f_6f_2^*), ~
\gamma_{26}=-\frac{1}{2}Re(f_4f_5^*+f_3f_6^*),~
\nn\\
\gamma_{27}&=&-\frac{1}{2}Re(f_1f_4^*+f_2f_3^*),  
\gamma_{28}=\frac{1}{2}Im(f_4f_5^*+f_3f_6^*),~
\gamma_{29}=-\frac{1}{2}Im(f_1f_4^*+f_2f_3^*), ~
\nn\\
\gamma_{30}&=&-\frac{1}{2}Re(f_4f_5^*-f_3f_6^*),  
 \gamma_{31}=-\frac{1}{2}Re(f_1f_4^*-f_2f_3^*),~
\gamma_{32}=-\frac{1}{2}Im(f_3f_6^*-f_4f_5^*),~
\nn\\
\gamma_{33}&=&-\frac{1}{2}Im(f_1f_4^*-f_2f_3^*),  
\gamma_{34}=-\frac{1}{2}Re(f_4f_6^*+f_3f_5^*),~
\gamma_{35}=-\frac{1}{2}Re(f_1f_3^*+f_2f_4^*),~
\nn\\
\gamma_{36}&=&\frac{1}{2}Im(f_3f_5^*+f_4f_6^*),  
\gamma_{37}=-\frac{1}{2}Im(f_1f_3^*+f_2f_4^*),~
\gamma_{38}=-\frac{1}{2}Re(f_4f_6^*-f_3f_5^*),~
\nn\\
\gamma_{39}&=&-\frac{1}{2}Re(f_2f_4^*-f_1f_3^*),  
\gamma_{40}=-\frac{1}{2}Im(f_3f_5^*-f_4f_6^*),~
\gamma_{41}=-\frac{1}{2}Im(f_2f_4^*-f_1f_3^*).
\label{eq:eqgamma}
\ea
The structure of the hadronic tensor describing the polarization of
the antiproton is the same as for the case of polarized proton. Let
us designate these structure functions as $\bar\beta_i$ and their
expressions in terms of the scalar amplitudes are
\ba \bar\beta_1&=&-2Imf_5f_6^*,
 ~\bar\beta_2=-2Imf_1f_2^*,~
\bar\beta_3=2Imf_3f_4^*,~ 
\bar\beta_4=Im(f_2f_5^*-f_1f_6^*),~
 \nn\\
\bar\beta_5&=&Re(f_2f_5^*-f_1f_6^*), 
\bar\beta_6=Im(f_3f_6^*-f_4f_5^*),
~\bar\beta_7=Im(f_1f_4^*-f_2f_3^*),~
 \nn\\
\bar\beta_8&=&Re(f_3f_6^*-f_4f_5^*),~
\bar\beta_9=Re(f_2f_3^*-f_1f_4^*),
\bar\beta_{10}=Im(f_4f_6^*+f_3f_5^*),~
 \nn\\
\bar\beta_{11}&=&-Im(f_1f_3^*+f_2f_4^*),~
\bar\beta_{12}=Re(f_4f_6^*+f_3f_5^*),~
\bar\beta_{13}=Re(f_2f_4^*+f_1f_3^*),
\label{eq:eqbeta}
\nn
\ea
which differ from Eq. (\ref{eq:betai})  mainly by signs.

\end{document}